# Current-Controlled Skyrmionic Diode

Xiangjun Xing,[1,2] Philip W. T. Pong,[1] and Yan Zhou[3,4,*]

[1] *Department of Electrical and Electronic Engineering, The University of Hong Kong, Hong Kong, China.*

[2] *College of Physics and Electronic Information Engineering, Wenzhou University, Wenzhou 325035, China.*

[3] *Department of Physics, The University of Hong Kong, Hong Kong, China.*

[4] *School of Electronic Science and Engineering and Collaborative Innovation Center of Advanced Microstructures, Nanjing University, Nanjing 210093, China.*

**ABSTRACT**

In order to address many of the challenges and bottlenecks currently experienced by traditional charge based technologies, various alternatives are being actively explored to provide potential solutions of device miniaturization and scaling in the more-than-MOORE era. Amongst these alternatives, spintronics physics and devices have recently attracted a rapidly increasing interest by exploiting the additional degree of electron's spins. For example, magnetic domain-wall racetrack-memory and logic devices have been realized via manipulating domain-wall motion. As compared to domain-wall based devices, magnetic skyrmions have the advantages of ultra-small size (typically 5–100 nm in diameter), facile current-driven motion, topological stability and peculiar emergent electrodynamics, promising for next-generation electronics applications in the more-than-Moore regime. In this work, a magnetic meron diode, which behaves like a PN-junction diode, is demonstrated for the first time, by tailoring the current-controlled unidirectional motion of edge-merons (i.e., fractional skyrmions) in a nanotrack with interfacial Dzyaloshinskii-Moriya interaction. The working principles of the meron diode, theoretically expected from the Thiele equation for topological magnetic objects, are further verified by micromagnetic simulations. The present study reveals topology-independent transport property of magnetic objects, and is expected to open the vista toward integrated composite circuitry, with unified data storage and processing, based on a single magnetic chip, as the meron diode can be used either, as a building block, to develop complex logic components or, as a signal controller, to interconnect skyrmion, domain-wall, and even spin-wave devices.

---
[*]Corresponding author. E-mail: yanzhou@hku.hk

## INTRODUCTION

Magnetic skyrmions are topologically stable configuration of magnetization vectors with a fixed skyrmion number ($Q$), which exists in non-centrosymmetric bulk magnets (1, 2) or ultrathin magnetic multilayer films lacking inversion symmetry (3), where asymmetric Dzyaloshinskii-Moriya interaction (DMI) (4–7), mediated by certain heavy-metal atoms with strong spin-orbit coupling, tends to twist the neighboring spins. The quantized topological charge protects a skyrmion against pinning by defect (8–10), but causes transverse displacement, during its drift motion along a transmission channel, by inducing a Magnus force once *e.g.* a current is applied, giving rise to the so-called skyrmion Hall effect (8, 9, 11, 12), which is undesirable in practical applications such as skyrmion-based racetrack memory, and can be eliminated by introducing a bilayer composite structure (13). At sufficiently large current densities, where the Magnus force overcomes the boundary's repulsion force, a skyrmion is pushed to touch the lateral boundary of the transmission track, forming a fractional skyrmion (termed *edge-meron* (14)). Then, the inward directed repulsion force on the skyrmion converts into an outward directed attraction force upon the edge-meron. In terms of string geometry (15), an edge-meron is enclosed by a curved open string and a lateral boundary of the transmission channel, and, therefore, it can be deemed as an intermediate spin texture between a skyrmion with $Q$=1 and a domain-wall pair with $Q$=0. Intrinsically, edge-merons are highly unstable instantons (16) for loss of topological protection. Without current applied, they will decay rapidly. According to the Thiele equation (17), the edge-meron would experience a Magnus force (18–20) once a current is employed, since the skyrmion number for an edge-meron is still finite despite being smaller than 1 (14). The direction of the Magnus force depends on the direction of the in-plane current injected into the transmission track (8–10). Thus, it is possible to tune the Magnus force to make it favor or react against the boundary's attraction force by changing the current direction.

By means of micromagnetic simulations, we address the current-driven dynamics of magnetic edge-merons in a nanotrack made of an ultrathin multilayer film exhibiting interfacial DMI (6, 7). We find that, for a certain current direction, the Magnus force on the edge-meron can indeed counteract the boundary's attraction force, resulting in its dynamical stabilization in the transverse direction and steady flow along the lateral-boundary channel with a velocity proportional to the applied current density. When the current direction is reversed, the edge-meron is repelled from the boundary. These findings are in accordance with the prediction of the Thiele equation (18, 19). Remarkably, the mobility, *i.e.*, the velocity over the current density, for the edge-merons and skyrmions, appears to be totally unrelated to the topological charges of these magnetic objects as well as the material parameters, even though the topological charge of an edge-meron evolves with the applied current density.

The fact that the nonreciprocal motions of an edge-meron along the boundary channel is dependent on the current direction promises a *current-controlled meron diode*, which is patterned into a lateral junction with a wide nanotrack serving as the modulation element and two narrow ones as the output element. The in-plane current is chosen as the control signal (playing a role as the bias voltage in conventional PN-junction diode (21)), and the inductive voltage in the detection coil traversed by a moving domain wall (converted from an edge-meron stabilized in the central, modulation track) is encoded into the output signal. Numerical simulations demonstrate that the proposed diode can operate over a broad range of parameter space and even at room temperature. Signal processing, based on the meron diode, together with the well-known domain-wall logic (22) and racetrack-memory (23) technologies, should lay the foundation for development of magnetic computers (24, 25) beyond the von Neumann architecture with strictly separated logic and memory.

**RESULTS**

**Theoretical prediction based on the Thiele equation**

The Landau-Lifshitz-Gilbert equation is a well-established general-purpose tool in describing spin dynamics of any continuous ferromagnetic system (8–10, 13, 15, 18, 19, 24, 26, 27). From this general equation of motion of magnetization, the so-called Thiele equation can be obtained to describe the motion of center of mass of a rigid spin texture (8, 10, 19),

$$\mathbf{G} \times (\mathbf{v}^s - \mathbf{v}^d) - \nabla V + \mathcal{D}(\beta \mathbf{v}^s - \alpha \mathbf{v}^d) = 0 \qquad (1)$$

expressing the balance of the Magnus, confining, and viscous forces, where the gyrocoupling vector $\mathbf{G} = G\hat{e}_z$ with $\hat{e}_z$ representing the unit vector along the vertical $z$ axis, $V$ is the confining potential due to boundaries, impurities, and/or magnetic fields, and $\mathcal{D} = \begin{pmatrix} \mathcal{D}_{xx} & \mathcal{D}_{xy} \\ \mathcal{D}_{yx} & \mathcal{D}_{yy} \end{pmatrix} = \begin{pmatrix} \mathcal{D} & 0 \\ 0 & \mathcal{D} \end{pmatrix}$ is a dissipation tensor. $\alpha$ is the Gilbert damping parameter, $\beta$ is the relative strength of the nonadiabatic and adiabatic spin torques in the Zhang-Li form associated with an in-plane current (28, 29), $\mathbf{v}^d$ is the drift velocity of the spin texture, and $\mathbf{v}^s$ is the velocity of conduction electrons, which is equal to the electron current density $\mathbf{j} = -J\hat{e}_x$ multiplied by a prefactor $\gamma\hbar P/(2\mu_0 e M_s)$, where $J$ is the magnitude of electric current density, $\hat{e}_x$ the unit vector in the $x$ direction, $\gamma$ the gyromagnetic ratio, $\mu_0$ the vacuum permeability, $\hbar$ the reduced Planck constant, $P$ the spin polarization of flowing electrons in the nanotrack, $e$ the elementary charge, and $M_s$ the saturation magnetization.

The gyroconstant $G$ is proportional to $Q$ (8, 9), which is 1 for a skyrmion and 0 for a domain wall. Consequently, the Magnus force $\mathbf{F}_g = \mathbf{G} \times (\mathbf{v}^s - \mathbf{v}^d)$ will act on a moving skyrmion if $\mathbf{v}^d \neq \mathbf{v}^s$, and it is always absent for domain walls. When a skyrmion moves along a nanotrack, it experiences bilateral confining potential; therefore, once the skyrmion, under the Magnus force, departs from the center of the track to approach one of the two borders, the confining force $\mathbf{F}_p = -\nabla V$ will emerge from that border as an opposing force (8, 9). Finally, under appropriate driving current densities, the skyrmion will remain stabilized transversally and drift steadily along the nanotrack (8, 9), that is, the Magnus force can always be compensated by the confining force from either boundary, irrespective of the skyrmion's drift direction.

Provided that an edge-meron can preserve a rigid structure, the current-driven motion of edge-merons should satisfy the above Thiele equation. The edge-meron lies at a specific border; thus, the force $\mathbf{F}_p$ from that border has a definite direction. As a result, if the force $\mathbf{F}_p$ is oppositely directed with respective to $\mathbf{F}_g$ for a certain current direction, they will point in the same direction when the applied current is reversed (as illustrated in Fig. 1). Equation (1) requires that the topological charge $Q$ is nonzero and $\mathbf{v}^d \neq \mathbf{v}^s$ in order for a finite $\mathbf{F}_g$ on the edge-meron to occur. By applying an in-plane current along the track (*i.e.* the $x$ axis) and assuming that, under the given current density, the edge-meron reaches motional steady state, one gets $v_y^s=0$, $v_y^d=0$, $\mathcal{D}(\beta \mathbf{v}^s - \alpha \mathbf{v}^d)=0$, and $\mathbf{F}_g+\mathbf{F}_p=0$. After some calculation, one obtains (9)

$$v_x^d = (\beta/\alpha) v_x^s \tag{2}$$

$$\text{and } F_g = (1-\beta/\alpha) G v_x^s = -F_p \tag{3}$$

which requires $1-\beta/\alpha \neq 0$, $G \neq 0$, and $v_x^s \neq 0$ (*i.e.*, $\beta \neq \alpha$, $Q \neq 0$, and $J \neq 0$) for an edge-meron to enter into steady drift motion. Otherwise, the edge-meron will destabilize and annihilate finally. It is worthy to note that $J$ is an adjustable parameter, easily tunable in terms of its amplitude and direction, in tailoring the alignment of forces. As a next step, we resort to micromagnetic simulations to test the assumptions and theoretical predictions made herein.

**Numerical verification based on micromagnetic simulations**

The nanotracks, used as transmission channel of magnetic merons, are patterned from an ultrathin multilayer film, with asymmetric interfaces to engender an interfacial DMI (6, 7). In what follows, we will demonstrate, firstly, the fundamental principle of nonreciprocal edge-meron transport, along the boundary channel, driven by in-plane electric currents. Here, we use a magnetic nanotrack with a length of 1200 nm and a width of 60 nm, in which an edge-meron is preset and then moved by an in-plane current (Figs. 2–4 and Supplemental figs. S1–S2). Current-driven skyrmion motion is also examined in the same nanotrack for comparison (8, 9). Subsequently, we will check the influence of edge irregularity on meron motion. For this, a triangular notch (8, 30) with variable depth is included into the border of the nanotrack to mimic the boundary defect (31) (Fig. 5). Finally, we will demonstrate how a meron diode works by virtue of the current-modulated unidirectional motion of edge-merons. To this end, a planar-junction-type structure, composed of nanotracks with different widths (15, 32), is adopted (Fig. 6 and Supplemental figs. S3–S4). The thickness of the tracks for all the simulations is 1 nm.

**Unidirectional motion of edge-merons**

The rigidity of the meron spin configuration is well maintained during its motion inside a track without imperfections, as clearly seen from Fig. 2 (A and C), where the edge-meron moves smoothly, showing stable shape and structure, especially after the establishment of steady drift motion characterized by unvaried $m_z$ and $Q$ with time after 1.5 ns [Fig. 2 (B and D)]. The steady drift of the edge-meron along the track [Fig. 2 (A and C)] implies that the Magnus force occurs to the meron and offsets the drag force of the boundary. Without the current-induced Magnus force, the meron will be pulled out of the track soon (fig. S1C), where $\mathbf{v}^d=\mathbf{v}^s$ (resulting from $\alpha=\beta$) permits no gyrotropic force, as expected from Eq. (3). The existence of the Magnus force on the meron suggests that the edge-meron has nonzero topological charge, as is confirmed by the numerical values of $Q$ [Fig. 2 (B and D)] directly calculated from the simulated spin configuration according to $Q=(1/4\pi)\int\mathbf{m}\cdot(\partial_x\mathbf{m}\times\partial_y\mathbf{m})dxdy$ (15).

By reversing the current direction in Fig. 2A but keeping the other parameters unaltered, we arrive at the results in Fig. 2 (E and F) displaying that the edge-meron decays quickly. Apparently, the Magnus force and the attraction force from the boundary combine into an outward net force, which drags the meron. That is to say, the Magnus force can be reversed by simply reversing the current direction, verifying the theoretical prediction [Eq. (3)] of the Thiele equation. The orientation of the Magnus force determines the dynamics of the edge-meron, as seen from comparing Fig. 2 (A and B) and Fig. 2 (E and F). In a word, when the Magnus force balances the boundary's attraction force in the transverse direction, the meron drifts steadily along the boundary channel; when the Magnus force is opposite to the boundary's force, the meron is annihilated, soon after injection, at a timescale of hundreds of picoseconds [Fig. 2 (E and F) and fig. S1 (A–C)]. In this way, unidirectional transmission of the meron carriers is realized.

In Fig. 2C, $1-\beta/\alpha=0.5$ has a sign opposite to $1-\beta/\alpha=-1$ in Fig. 2A, and meanwhile the current directions are also opposite there. Considering that the boundary's force is always outward directed, the directions of the Magnus forces in Fig. 2 (A and C) must be identical and inward directed to keep the steady-state motion. This agrees with the anticipation of Eq. (3). At this point, Eq. (3) (the dependencies of $\mathbf{F}_g$ on $\beta/\alpha$, $Q$, and $J$) has been thoroughly substantiated by simulation results.

According to Eq. (2), the velocity of a spin texture has nothing to do with its topological charge, which is revealed in Fig. 2C, where a skyrmion with $Q\sim1$ and an edge-meron with $Q\sim0.5$ move synchronously along the track just as bound together, despite no interaction between them. This independency will be further validated by additional simulation results in Fig. 4 indicating the topological charge of an edge-meron changing with the driving current density. The good agreement between the simulated and analytical results suggests that the meron dynamics under in-plane currents can be well captured by Eq. (1)—the massless Thiele equation—at least for the case of perfect tracks without defects.

To reach the motional steady state, the drag force from the boundary must be rigorously offset by the Magnus force. The boundary's force the edge-meron senses is determined by the potential landscape of the track (8, 9), which is related to the material parameters and the shape, size, and topological charge of the meron. The topological charge of the edge-meron exhibits dependency on the driving current density. It is impossible to derive an explicit expression for $V=V[K_u, D, Q(K_u, D, J)]$ and thus for $\mathbf{F}_p=-\nabla V$. In turn, the current-density window guaranteeing steady meron motion cannot be analytically extracted, and numerical simulations become a proper tool to address such issue. We examine the current-driven motion of an edge-meron in the nanotrack, with $K_u$ varying from 0.4 to 1.2 and $D$ varying from 2.0 to 4.5 covering the range of the most technological relevance (8, 15, 24, 27, 33, 34), beyond which a regular skyrmion is not allowed to exist in the track in the remnant state [single-domain configuration for lower ($K_u$, $D$) (8); elongated-skyrmion or multi-domain configuration for higher ($K_u$, $D$) (8, 15)]. The results are presented in Fig. 3 as a phase diagram. The colored interior of the ring (named *stabilization ring*) centered at each ($K_u$, $D$) stands for the range of the current densities, under which the steady meron motion can be established. The inner area surrounded by the colored ring groups such current densities, at which the Magnus force is not large enough to compensate the boundary's force so that the meron is repelled from the track, whereas for the region outside each ring, the current density deforms an edge-meron into a domain-wall pair, by inducing a much stronger Magnus force than the oppositely directed force from the boundaries.

The stabilization rings are not identical for various material parameters. For a given $K_u$ with $α=0.3$ and $β=2α$, the higher the $D$ value, the larger is the outer radius of the ring and the wider the ring (Fig. 3A). At a given ($K_u$, $D$), the stabilization ring for ($α$, $β$)=(0.3, 0.5$α$) is wider and bigger than for ($α$, $β$)=(0.3, 2$α$) [Fig. 3 (B and C)]. The difference in stabilization rings reflects the complex reliance of the Magnus and boundary's forces upon the material properties. Specifically, the two forces are directly associated with the material parameters as well as the geometrical and/or topological characteristics (size, shape, topological charge etc.) of the edge-meron, as revealed in Eq. (3). The geometry and topology of a meron are also dependent on material properties. Associating the above considerations with Eq. (3), one can gets $J_s ∝ -∇V[K_u, D, Q(K_u, D, J_s)]/Q(K_u, D, J_s)$, where $J_s$ are the current densities allowing steady meron motion to be established, provided that other material parameters are given. The above implicit function of $J_s$ reveals the difficulty in analytically deriving the stabilization phase diagram and the dependence of the stabilization rings on material parameters.

Figure 4A shows the simulated velocity ($v_x^d$) versus current density ($J$) for steadily moving edge-merons in nanotracks with varied $K_u$, $D$, $α$, and $β$ values. It is clear that the drift velocities of edge-merons are linearly proportional to the driving current densities (8, 9), which is consistent with the expectation of Eq. (2). Moreover, defining the mobility of edge-merons as the velocity divided by the driving current density, *i.e.*, $v_x^d/J$, one can see that the mobility is independent of $K_u$, $D$ (material parameters) and $Q$ (topology parameter), as long as the current density is within the corresponding stabilization ring (Fig. 3), although the topological charge $Q$ changes with the current density $J$ [Fig. 4 (B–G)]. Note that, for the relevant current densities, the $Q$ value of an edge-meron is in the range of 0.4–0.65, which is not far from 0.5 (14). Once $Q$ becomes too large or too small, the force balance on the meron will be broken immediately [recall that, the Magnus force $F_g \propto J \times Q(J)$], and the meron will in turn collapse into a domain-wall pair (15) (fig. S1D) or disappear [Fig. 2 (E and F) and fig. S1 (A–C)]. Intriguingly, it appears that the skyrmions and edge-merons have the same mobility, when identical $\alpha$ and $\beta$ values are assumed in simulations, which evidences that the mobility of a spin texture in a given track is not affected by its topological charge, if the structural rigidity is well maintained during its motion. The observation that the mobility is independent of $K_u$, $D$ and $Q$ is in line with Eq. (2), where these parameters are absent and not implicitly involved as well.

According to Eq. (2), the meron mobility $\mu \equiv v_x^d/J = [\gamma\hbar P/(2\mu_0 e M_s)] \cdot (\beta/\alpha)$. Substituting the values of all constants and some parameters into the above formula, one gets $\mu = (0.400 \times 10^{-10} \beta/\alpha)$ m$^3$A$^{-1}$s$^{-1}$. Thus, the theoretical mobility values are $0.800 \times 10^{-10}$ m$^3$A$^{-1}$s$^{-1}$ for $\beta = 2\alpha$ and $0.200 \times 10^{-10}$ m$^3$A$^{-1}$s$^{-1}$ for $\beta = 0.5\alpha$. From Fig. 4A, one finds that, at $\alpha = 0.3$, the simulated mobility values are $0.583 \times 10^{-10}$ m$^3$A$^{-1}$s$^{-1}$ for $\beta = 2\alpha$ and $0.194 \times 10^{-10}$ m$^3$A$^{-1}$s$^{-1}$ for $\beta = 0.5\alpha$; whereas at $\alpha = 0.01$, the values are

$0.583×10^{-10}$ m$^3$A$^{-1}$s$^{-1}$ for $β=2α$ and $0.381×10^{-10}$ m$^3$A$^{-1}$s$^{-1}$ for $β=0.5α$. The clear dependency of the mobility upon the damping parameter, observed in simulation results, is missing from the theoretical prediction. These slight discrepancies between the theory and simulations should be ascribed to the incompleteness of the massless Thiele equation in the rigid-body picture (8, 10, 19, 20), which neglects the relaxation of the internal spins and the structural deformation of a spin texture strongly relying on the damping properties of materials.

**Effect of boundary roughness**

In deriving the theoretical velocity and force equations, we assumed an ideal nanotrack without including any impurity or edge roughness. However, experimentally, a nanotrack, prepared even by the state-of-the-art microfabrication techniques, cannot avoid defects, such as boundary irregularity, which will affect the motion of spin textures in the track (31). As argued above, an edge-meron is an intermediate thing between a skyrmion and a domain-wall pair. It should behave like a skyrmion in the interior and like a domain-wall pair on the border line of a track. To clarify how boundary defect influences the motional dynamics of an edge-meron under an electric current, we introduce a triangular notch (8, 30) into the border of a nanotrack in simulations (inset of Fig. 5A). We find that the behavior of a meron in passing through the notch depends on the depth of the notch and the current density (8) [compare Fig. 5 (A and B)]. For instance, at a given current density of $J=+4.0×10^{12}$ Am$^{-2}$, the meron can pass a notch 3 nm in depth (5% of the track width) but cannot traverse a notch 6 nm in depth. If the current density is increased to $J=+6.0×10^{12}$ Am$^{-2}$, the meron can overcome all notches with depths of 3, 6, and 12 nm and return to the original trajectory.

As shown in Fig. 5A, during the motion of the meron toward the right end, the front wall meets the notch and then detaches, and a moment later the back wall touches the notch but is tightly pinned instead. The reason why the front wall can escape from the notch is that it senses the joint forces of the current and the back wall. The current exerts a viscous force via spin transfer; the back wall imposes a repulsive force through exchange interaction (35). By contrast, the back wall only experiences the viscous force due to the current, since the front wall is driven away from the back wall and thus cannot offer a force (even though the front wall is close to the back wall, it cannot help the latter to depin from the notch, because its repulsive force counteracts the viscous force of the current). In pushing the front wall forward, the current elongates the meron into a strip domain (15). However, the picture is different for the meron under a higher current density, as shown in Fig. 5B, where the viscous force from the current is so large that both the front and back walls can easily escape from the notch.

**On symmetry breaking**

Figures 2–3 and figs. S1 and S2 contain solid evidence to support the prediction (based on the Thiele equation) that the motions of edge-merons, under in-plane currents, are nonreciprocal, over a wide range of space of material parameters $K_u$, $D$, $\alpha$, and $\beta$. Intrinsically, the unidirectionality in the meron motion should stem from the breaking in the mirror symmetry of the potential landscape [viz. $V(-y) \neq V(y)$] of the nanotrack, where the spin texture is attached to one of the two symmetric lateral boundaries. Such a potential environment makes the boundary's confining force on the meron to be locked into a specific orientation and to be unable to balance the Magnus force for one of the two current directions, leading finally to the current-controlled unidirectional motion of the edge-merons. The nonreciprocal meron motion, benefiting from special characteristics of 'edge states', bears some resemblance to the edge-localized propagation of the Damon-Eshbach spin waves in a 1-dimensional magnetic waveguide (32, 36, 37). The spin-wave edge channels—the potential wells (minimums in the internal field)—near the lateral boundaries of a waveguide are induced by the boundary magnetic charges (38, 39), which can be created only if the translational symmetry of the waveguide is broken in its width direction. The occurrence of the spin-wave edge states, by introducing spatially separated edge channels to accompany the original center channels, makes possible spin-wave confluence and beating in a single waveguide (40), which could find potential application in multichannel information transmission and processing and nanometer-scale frequency deconvolution of microwave signals (36).

**Meron-based diode**

We propose a magnetic meron diode (Fig. 6A), the key element of which is a lateral junction consisting of a wide track and two narrow arms. An edge-meron is injected into the wide track by the Slonczewski spin torque of a perpendicular current, which is applied to a local area covered by a point-contact spin valve (8, 41). To manipulate the meron, an in-plane current will be fed into the junction, using a connected control unit, immediately after the termination of the injection current. The detection circuit records a signal, once a domain-wall pair passes through a coil.

The operation processes of the diode are as follows. For both forward and reverse cycles, six repeated operations are implemented sequentially. In each cycle, the perpendicular current ($J_1$) is firstly used and then the in-plane current ($J_2$) [Fig. 6 (B and E)]. Fig. 6D addresses the forward process. After nucleated, a meron is pushed to move rightward and later converted into a domain-wall pair at the interface between the wide and narrow tracks (15). When the domain-wall pair goes through the section beneath the coil, the latter senses a varying magnetic flux and produces an electromotive force. Finally, the domain-wall pair leaves the junction from the right terminal. More than one merons can proceed in the junction simultaneously; there is no coupling between any two of the merons and domain-wall pairs (8), if the temporal profile of the current sequence is well designed. The duration of $J_1$ cannot be too short in order for a meron to be formed [the injection processes of an edge-meron are illustrated in fig. S3 for several sets of ($K_u$, $D$, $\alpha$)]. $J_2$ should be sufficiently long to prevent clogging of merons in the track (15). As shown in Fig. 6C, the vertical magnetization decreases with the increase in the number of merons injected into the junction. At 0.80 ns, the domain-wall pair touches the right end of the junction, and, the vertical magnetization begins to rise. The periodic oscillation of magnetization features the reproducible manipulation of merons by the repeated current pulses. As Fig. 6G indicates, for the reverse process, only a single meron is present in the junction at a given time; that is because the former meron has been dissolved during the action of each $J_2$, not until the initiation of the next $J_1$. In this case, the merons are annihilated in the central track and cannot enter the narrow arms to contribute an electromotive voltage.

Quantifying the forward and reverse processes, one can acquire the characteristic curve of the meron diode as shown in Fig. 6H (equivalent to the '*I–V*' curve of a PN-junction diode (21)). The output of the diode is encoded as the electromotive force, $\varepsilon$, induced in the coils of the detection circuit. After some calculation, one can arrive at $\varepsilon \equiv -d\Phi/dt = 2\mu_0 M_s \cdot w_n \cdot v_{dw}$, where $\Phi$ is the total magnetic flux across the coils (here, we assume a tiny spacing between the coil and junction planes and $\mathbf{B}=B\hat{\mathbf{e}}_z$ and $B \approx \mu_0 H_z^d \approx \mu_0 M_s$, with $\mathbf{B}$ and $H_z^d$ the magnetic induction and the vertical component of the stray field from the narrow track at a point of the coil plane, respectively), $w_n$ the width of the narrow arms, and $v_{dw}$ the domain-wall velocity in the narrow arm. Finding $v_{dw}(J)$ using micromagnetic simulations and substituting it into the above expression, the $\varepsilon(J)$ is identified. On the reverse side, since the merons cannot reach the coils, the output is always zero. On the forward side, there exists a threshold current density ($J_{c1}$), below which the meron cannot be sent into the narrow arm, resulting in an empty output. In fact, there is another threshold current density ($J_{c2}$, corresponding to the smallest periphery of a ring in Fig. 3) below which the steady meron motion is not permitted. However, the $J_{c2}$ is smaller than $J_{c1}$ and thus unable to manifest itself in the $\varepsilon$–$J$ curve. Above $J_{c1}$, the output, $\varepsilon$, is directly proportional to the driving current density, $J$, because $\varepsilon \propto v_{dw}$ and in turn $v_{dw} \propto J$ [as is known from simulations and Eq. (2)]. The driving current cannot be too large; otherwise, excessive spin textures will nucleate at the ends of the junction (42) and move against the electric current, which will totally disrupt the regular operation of the device. Moreover, a high current might cause strong chaoticity in spin dynamics (42, 43) and even damage the sample by generating the Joule heating. The finite-temperature micromagnetic simulations reveal that the diode can work at room temperature (fig. S4). The junction might not be overheated by the Joule heating, as the current pulses in each operation cycle are sufficiently short, lasting for 320 ps in the demonstrated case. In real devices, the interval between cycles should be optimized to allow cooling the sample via thermal dissipation.

As noticed from simulations, the domain-wall pair becomes wider as the current density increases (Fig. 6I). The explanation is as follows. The width of the domain-wall pair is determined by the domain-wall velocity in the narrow track and the time required for a meron to be converted into a domain-wall pair. Both the domain-wall velocity and the conversion time are functions of the current density.

In Fig. 6 (B and E), the current $J_2$ is pulsed and applied after $J_1$. In fact, our simulations demonstrate that $J_2$ can be utilized continuously (as a direct current) and only $J_1$ needs to be pulsed to periodically inject merons. Of course, the use of a direct current is not a good choice, from the point of view of heat dissipation.

The driving current, $J$, in the junction comes from an external voltage, $U$, supplied by the control circuit. Substituting $U(J)$ into $\varepsilon(J)$, one finds that $\varepsilon/U=(\gamma\hbar/e)(\beta/\alpha)P\cdot\sigma\cdot[1/(l_b/w_b+l_n/w_n)]$, where $\sigma$ is the conductivity of the junction material, and $l_b$ ($l_n$) and $w_b$ ($w_n$) are the length and width of the wide (narrow) track in the junction, respectively. This means that the ratio of the output to input voltages is independent of the current density and instead is determined by the geometric ($l_b$, $w_b$, $l_n$, and $w_n$) and material ($\beta$, $\alpha$, $P$, and $\sigma$) parameters of the junction. As such, materials with higher $\beta/\alpha$ (44, 45) will bring enhanced output signal, at a given current density, or lowered current-density range, at a given magnitude of the output signal [as $\varepsilon \propto v_{dw} \propto (\beta/\alpha)J$]. It should be emphasized that the operation process of the diode depends heavily on $J$, the driving current density.

For $\alpha=\beta$, the forward and reverse motions of the meron are equivalent, that is, no nonreciprocity is associated with the meron motions. Because of the absence of the Magnus force, the meron moves along the electron current and meanwhile decays under the outward directed drag force from the boundary. Consequently, the diode cannot work at $\alpha=\beta$.

To prevent the merons from entering into the left arm, the central track should be made longer than the distance (tens to hundreds of nanometers; see fig. S2), which a meron travels, from its injection to annihilation.

**DISCUSSION**

The Thiele equation, Eq. (1), neglects the mass of the moving merons, and thus can only approximately uncover the real dynamics of the current-driven merons, which causes slight quantitative discrepancies between the theoretical prediction and the simulation results. However, the key prediction of the Thiele equation—the unidirectional motion of the edge-merons—is confirmed by micromagnetic simulations. Therefore, the massless Thiele equation (8, 10, 19, 20) captures the core element of the meron motion dynamics in this system. The generalized Thiele equation (46, 47), considering the mass of the merons, can be developed to improve understanding on the meron dynamics, which is however beyond the scope of the present paper.

Recently, reliable conversion between a skyrmion and a domain-wall pair has been demonstrated (15, 48), and multiple interaction schemes between domain walls/skyrmions and spin waves, have been identified (24, 27, 34, 41, 49–53). Besides the well-known fact that domain walls are capable of modulating the propagation characteristics of spin waves (49, 50), it was demonstrated most recently that a magnetic nanotrack, with imprinted domain-wall lines, can serve as a graded-index 'optic fiber' for channeling spin waves (24, 27, 34). On contrary, propagating spin waves can trigger domain-wall/skyrmion motion via a magnonic spin-transfer torque (52, 53). These findings enrich the family of magnetic logic and memory devices (15, 22, 23, 27, 49, 54, 55). The diode presented here is built on a planar, track-based structure, which has been adopted in both domain-wall/skyrmion logic and racetrack memory devices (15, 22, 23, 27, 49), and, remarkably, has also been utilized in the mainstream magnonic logic devices based on propagating spin waves (54, 55). These facts imply that the meron diode can be directly integrated to the aforementioned logic and memory circuits as a signal controller, and, furthermore, can be conveniently reconfigured to perform other functions (8, 15, 27, 34) as a reprogrammable device. Thus, the magnetic meron diode, as a new member of the diode family (24, 56–61), is anticipated to play a crucial role in information processing and data storage based on the magnetic features—skyrmions, domain-walls, and even spin waves.

It is noteworthy to note that, in the proof-of-principle demonstration of the proposed diode, we use the Zhang-Li form of spin-transfer torque to drive edge-merons into motion, and practically, the device performance can be greatly enhanced by optimizing the used materials, device geometry, and driving schemes. Alternatively, the emergent spin-orbit torques (spin Hall torque and/or Rashba torque) should be chosen to move the edge-merons in real devices, because they might allow the diode to work under much reduced current densities, and additionally the restriction β α required for diode operation with the Zhang-Li torque can be released. In fact, most recently, our collaborators have experimentally observed the skyrmion Hall effect, for which the current-induced spin Hall torques were used to drive skyrmions into motion (62). This experiment gives a strong hint that the proposed diode should function practically and the spin-Hall torque should be an effective means for operating the meron diode.

**METHODS**

Micromagnetic simulations based on MuMax3 (63) are carried out to study the injection of an edge meron under a perpendicular current, and to trace the dynamics of meron motion driven by an in-plane current. For all computations, the interfacial DMI (64) is added into the conventional LLG equation (65, 66), for those computations examining spin dynamics triggered by the out-of-plane current, the Slonczewski spin torque (67) is included as well, and for those tackling spin dynamics stimulated by the in-plane current, the Zhang-Li spin torque (28, 29) is incorporated additionally. For finite-temperature simulations, the random thermal field of the Brown form (63) is included in the effective magnetic field (the results are shown in fig. S4). The material parameters typical of Pt/Co multilayer systems with perpendicular magnetic anisotropy are employed in simulations (8, 15): $M_s$=580 kAm$^{-1}$, the exchange stiffness $A$=15 pJm$^{-1}$, $P$=0.4, and $\alpha$=0.3 ($\alpha$=0.05 (9, 68) and 0.01 (33, 69) are also examined in simulations to see the influence of damping constant; see Fig. 4 and fig. S2). According to Eq. (3), $\beta/\alpha$=1 shall lead to zero Magnus force and thus destabilization of the edge-meron motion; therefore, the other two representative cases of $\beta/\alpha$=2 and 0.5 are considered in simulations. A series of $K_u$ (perpendicular magnetocrystalline anisotropy) and $D$ (the DMI strength) combinations were examined in computations to ensure the obtained results are valid for a variety of samples (8, 15) (Fig. 3 and Supplemental figs. S2 and S3). The results presented in the figures throughout the paper correspond to $K_u$=0.8 MJm$^{-3}$ (the effective uniaxial anisotropy $K_{eff}$=0.6 MJm$^{-3}$ as given by $K_{eff}=K_u-(1/2)\mu_0 M_s^2$) and $D$=3.5 mJm$^{-2}$ unless specified otherwise. The computational volume is divided into regular meshes of 1×1×1 nm$^3$ regardless of the sample size.

**ASSOCIATED CONTENT**

**Supporting Information**: The Supporting Information is available.

Figures S1−S4 providing details on the current-dependent motion of edge-merons, the material-parameters dependence of the injection and unidirectional motion of edge-merons, and the diode effect at finite temperature.

**AUTHOR INFORMATION**

**Corresponding Authors:** *E-mail: yanzhou@hku.hk

**Author Contributions:** X.J.X conceived the idea and performed micromagnetic simulations. Y.Z. coordinated and supervised the project. Y.Z. proposed to examine parameters dependence of the results. All authors discussed the results. X.J.X. wrote the manuscript.

**Notes:** The authors declare no competing financial interest.

**ACKNOWLEDGEMENTS:** Y.Z. acknowledges the support by National Natural Science Foundation of China (project No. 1157040329), the Seed Funding Program for Basic Research and Seed Funding Program for Applied Research from the HKU, ITF Tier 3 funding (ITS/171/13, ITS/203/14), the RGC-GRF under Grant HKU 17210014, and University Grants Committee of Hong Kong (Contract No. AoE/P-04/08). X.J.X. thanks the support by the Zhejiang Provincial Natural Science Foundation of China under Grant No. LY14A040006 and the National Natural Science Foundation of China under Grant No. 11104206.

**REFERENCES**

1. Mühlbauer, S.; Binz, B.; Jonietz, F.; Pfleiderer, C.; Rosch, A.; Neubauer, A.; Georgii, R.; Böni, P. Skyrmion lattice in a chiral magnet. *Science* **2009**, 323, 915–919.

2. Yu, X. Z.; Onose, Y.; Kanazawa, N.; Park, J. H.; Han, J. H.; Matsui, Y.; Nagaosa, N.; Tokura, Y. Real-space observation of a two-dimensional skyrmion crystal. *Nature* **2010**, 465, 901–904.

3. Heinze, S.; von Bergmann, K.; Menzel, M.; Brede, J.; Kubetzka, A.; Wiesendanger, R.; Bihlmayer, G.; Blugel, S. Spontaneous atomic-scale magnetic skyrmion lattice in two dimensions. *Nat. Phys.* **2011**, 7, 713–718.

4. Dzyaloshinskii, I. A thermodynamic theory of "weak" ferromagnetism of antiferromagnetics. *J. Phys. Chem. Solids* **1958**, 4, 241–255.

5. Moriya, T. Anisotropic superexchange interaction and weak ferromagnetism. *Phys. Rev.* **1960**, 120, 91–98.

6. Fert, A.; Levy, P. M. Role of anisotropic exchange interactions in determining the properties of spin-glasses. *Phys. Rev. Lett.* **1980**, 44, 1538–1541.

7. Fert, A. Magnetic and transport-properties of metallic multilayers. *Mater. Sci. Forum* **1990**, 59, 439–480.

8. Sampaio, J.; Cros, V.; Rohart, S.; Thiaville, A.; Fert, A. Nucleation, stability and current-induced motion of isolated magnetic skyrmions in nanostructures. *Nat. Nanotechnol.* **2013**, 8, 839–844.

9. Iwasaki, J.; Mochizuki, M.; Nagaosa, N. Current-induced skyrmion dynamics in constricted geometries. *Nat. Nanotechnol.* **2013**, 8, 742–747.

10. Iwasaki, J.; Mochizuki, M.; Nagaosa, N. Universal current-velocity relation of skyrmion motion in chiral magnets. *Nat. Commun.* **2013**, 4, 1463.

11. Zang, J.; Mostovoy, M.; Han, J. H.; Nagaosa, N. Dynamics of skyrmion crystals in metallic thin films. *Phys. Rev. Lett.* **2011**, 107, 136804.

12. Kong, L.; Zang, J. Dynamics of an insulating skyrmion under a temperature gradient. *Phys. Rev. Lett.* **2013**, 111, 067203.

13. Zhang, X.; Zhou, Y.; Ezawa, M. Magnetic bilayer-skyrmions without skyrmion Hall effect. *Nat. Commun.* **2016**, 7, 10293.

14. Pereiro, M.; Yudin, D.; Chico, J.; Etz, C.; Eriksson, O.; Bergman, A. Topological excitations in a kagome magnet. *Nat. Commun.* **2015**, 5, 4815.

15. Zhou, Y.; Ezawa, M. A reversible conversion between a skyrmion and a domain-wall pair in a junction geometry. *Nat. Commun.* **2014**, 5, 4652.

16. Rajaraman, R. Solitons and Instantons, Volume 15: An Introduction to Solitons and Instantons in Quantum Field Theory (North-Holland, Amsterdam, 1987).

17. Thiele, A. A. Steady-state motion of magnetic domains. *Phys. Rev. Lett.* **1973**, 30, 230–233.

18. Yu, X. Z.; Kanazawa, N.; Zhang, W. Z.; Nagai, T.; Hara, T.; Kimoto, K.; Matsui, Y.; Onose, Y.; Tokura, Y. Skyrmion flow near room temperature in an ultralow current density. *Nat. Commun.* **2012**, 3, 988.

19. Iwasaki, J.; Koshibae, W.; Nagaosa, N. Colossal spin transfer torque effect on skyrmion along the edge. *Nano Lett.* **2014**, 14, 4432–4437.

20. Lin, S.-Z.; Reichhardt, C.; Batista, C. D.; Saxena, A. Particle model for skyrmions

in metallic chiral magnets: dynamics, pinning, and creep. *Phys. Rev. B* **2013**, 87, 214419.

21. Neudeck, G. W. The PN Junction Diode (Addison-Wesley, Reading, MA, 1989).

22. Allwood, D. A.; Xiong, G.; Faulkner, C. C.; Atkinson, D.; Petit, D.; Cowburn, R. P. Magnetic domain-wall logic. *Science* **2005**, 309, 1688–1692.

23. Parkin, S. S. P.; Hayashi, M.; Thomas, L. Magnetic domain-wall racetrack memory. *Science* **2008**, 320, 190–194.

24. Lan, J.; Yu, W.; Wu, R.; Xiao, J. Spin-wave diode. *Phys. Rev. X* **2015**, 5, 041049.

25. Dery, H.; Dalal, P.; Cywinski, L.; Sham, L. J. Spin-based logic in semiconductors for reconfigurable large-scale circuits. *Nature* **2007**, 447, 573–576.

26. Zhou, Y.; Iacocca, E.; Awad, A. A.; Dumas, R. K.; Zhang, F. C.; Braun, H. B.; Åkerman, J. Dynamically stabilized magnetic skyrmions. *Nat. Commun.* **2015**, 6, 8193.

27. Xing, X.; Zhou, Y. Fiber optics for spin waves. *NPG Asia Mater.* **2016**, 8, e246.

28. Zhang, S.; Li, Z. Roles of nonequilibrium conduction electrons on the magnetization dynamics of ferromagnets. *Phys. Rev. Lett.* **2004**, 93, 127204.

29. Thiaville, A.; Nakatani, Y.; Miltat, J.; Suzuki, Y. Micromagnetic understanding of current-driven domain wall motion in patterned nanowires. *Europhys. Lett.* **2005**, 69, 990.

30. Pushp, A.; Phung, T.; Rettner, C.; Hughes, B. P.; Yang, S-H.; Thomas, L.; Parkin, S. S. P. Domain wall trajectory determined by its fractional topological edge defects. *Nat. Phys.* **2013**, 9, 505–511.

31. Nakatani, Y.; Thiaville, A.; Miltat, J. Faster magnetic walls in rough wires. *Nat. Mater.* **2003**, 2, 521–523.

32. Demidov, V. E.; Jersch, J.; Demokritov, S. O.; Rott, K.; Krzysteczko, P.; Reiss, G. Transformation of propagating spin-wave modes in microscopic waveguides with variable width. *Phys. Rev. B* **2009**, 79, 054417.

33. Tomasello, R.; Martinez, E.; Zivieri, R.; Torres, L.; Carpentieri, M.; Finocchio, G. A strategy for the design of skyrmion racetrack memories. *Sci. Rep.* **2014**, 4, 6784.

34. Garcia-Sanchez, F.; Borys, P.; Soucaille, R.; Adam, J.-P.; Stamps, R. L.; Kim, J.-V. Narrow magnonic waveguides based on domain walls. *Phys. Rev. Lett.* **2015**, 114, 247206.

35. Zhang, X.; Zhao, G. P.; Fangohr, H.; Liu, J. P.; Xia, W. X.; Xia, J.; Morvan, F. J. Skyrmion-skyrmion and skyrmion-edge repulsions in skyrmion-based racetrack memory. *Sci. Rep.* **2015**, 5, 7643.

36. Demidov, V. E.; Demokritov, S. O.; Rott, K.; Krzysteczko, P.; Reiss, G. Nano-optics with spin waves at microwave frequencies. *Appl. Phys. Lett.* **2008**, 92, 232503.

37. Xing, X.; Li, S.; Huang, X.; Wang, Z. Engineering spin-wave channels in submicrometer magnonic waveguides. *AIP Adv.* **2013**, 3, 032144.

38. Jorzick, J.; Demokritov, S. O.; Hillebrands, B.; Berkov, D.; Gorn, N. L.; Guslienko, K.; Slavin, A. N. Spin wave wells in nonellipsoidal micrometer size magnetic elements. *Phys. Rev. Lett.* **2002**, 88, 047204.

39. Park, J. P.; Eames, P.; Engebretson, D. M.; Berezovsky, J.; Crowell, P. A. Spatially

resolved dynamics of localized spin-wave modes in ferromagnetic wires. *Phys. Rev. Lett.* **2002**, 89, 277201.

40. Demokritov, S. O.; Demidov, V. E. in Spin Wave Confinement, edited by Demokritov S. O. (Pan Stanford Publishing Pte. Ltd., Singapore, 2009), Chap. 1.

41. Ma, F.; Zhou, Y.; Braun, H. B.; Lew, W. S. Skyrmion-based dynamic magnonic crystal. *Nano Lett.* **2015**, 15, 4029−4036.

42. Sethi, P.; Murapaka, C.; Lim, G. J.; Lew, W. S. In-plane current induced domain wall nucleation and its stochasticity in perpendicular magnetic anisotropy Hall cross structures. *Appl. Phys. Lett.* **2015**, 107, 192401.

43. Seo, S. M.; Lee, K. J.; Yang, H.; Ono, T. Current-induced control of spin-wave attenuation. *Phys. Rev. Lett.* **2009**, 102, 147202.

44. Eltschka, M.; Wötzel, M.; Rhensius, J.; Krzyk, S.; Nowak, U.; Kläui, M.; Kasama, T.; Dunin-Borkowski, R. E.; Heyderman, L. J.; van Driel, H. J.; Duine, R. A. Nonadiabatic spin torque investigated using thermally activated magnetic domain wall dynamics. *Phys. Rev. Lett.* **2010**, 105, 056601.

45. Boulle, O.; Kimling, J.; Warnicke, P.; Kläui, M.; Rüdiger, U.; Malinowski, G.; Swagten, H. J. M.; Koopmans, B.; Ulysse, C.; Faini, G. Nonadiabatic spin transfer torque in high anisotropy magnetic nanowires with narrow domain walls. *Phys. Rev. Lett.* **2008**, 101, 216601.

46. Tretiakov, O. A.; Clarke, D.; Chern, G.-W.; Bazaliy, Ya. B.; Tchernyshyov, O. Dynamics of domain walls in magnetic nanostrips. *Phys. Rev. Lett.* **2008**, 100, 127204.

47. Makhfudz, I.; Krüger, B.; Tchernyshyov, O. Inertia and chiral edge modes of a skyrmion magnetic bubble. *Phys. Rev. Lett.* **2012**, 109, 217201.

48. Jiang, W.; Upadhyaya, P.; Zhang, W.; Yu, G.; Jungfleisch, M. B.; Fradin, F. Y.; Pearson, J. E.; Tserkovnyak, Y.; Wang, K. L.; Heinonen, O.; te Velthuis, S. G. E.; Hoffmann, A. Blowing magnetic skyrmion bubbles. *Science* **2015**, 349, 283–286.

49. Hertel, R.; Wulfekel, W.; Kirschner, J. Domain-wall induced phase shifts in spin waves. *Phys. Rev. Lett.* **2004**, 93, 257202.

50. Iwasaki, J.; Beekman, A. J.; Nagaosa, N. Theory of magnon-skyrmion scattering in chiral magnets. *Phys. Rev. B* **2014**, 89, 064412.

51. Hermsdoerfer, S. J.; Schultheiss, H.; Rausch, C.; Schafer, S.; Leven, B.; Kim, S.-K.; Hillebrands, B. A spin-wave frequency doubler by domain wall oscillation. *Appl. Phys. Lett.* **2009**, 94, 223510.

52. Han, D.-S.; Kim, S.-K.; Lee, J.-Y.; Hermsdoerfer, S. J.; Schultheiss, H.; Leven, B.; Hillebrands, B. Magnetic domain-wall motion by propagating spin waves. *Appl. Phys. Lett.* **2009**, 94, 112502.

53. Yan, P.; Wang, X. S.; Wang, X. R. All-magnonic spin-transfer torque and domain wall propagation. *Phys. Rev. Lett.* **2011**, 107, 177207.

54. Kostylev, M. P.; Serga, A. A.; Schneider, T.; Leven, B.; Hillebrands, B. Spin-wave logical gates. *Appl. Phys. Lett.* **2005**, 87, 153501.

55. Chumak, A. V.; Serga, A. A.; Hillebrands, B. Magnon transistor for all-magnon data processing. *Nat. Commun.* **2014**, 5, 4700.

56. Li, B.; Wang, L.; Casati, G. Thermal diode: rectification of heat flux. *Phys. Rev.*

*Lett.* **2004**, 93, 184301.

57. Chang, C. W.; Okawa, D.; Majumdar, A.; Zettl, A. Solid-state thermal rectifier. *Science* **2006**, 314, 1121–1124.

58. Liang, B.; Guo, X. S.; Tu, J.; Zhang, D.; Cheng, J. C. An acoustic rectifier. *Nat. Mater.* **2010**, 9, 989–992.

59. Wang, D.-W.; Zhou, H.-T.; Guo, M.-J.; Zhang, J.-X.; Evers, J.; Zhu, S.-Y. Optical diode made from a moving photonic crystal. *Phys. Rev. Lett.* **2013**, 110, 093901.

60. Borlenghi, S.; Wang, W.; Fangohr, H.; Bergqvist, L.; Delin, A. Designing a spin-Seebeck diode. *Phys. Rev. Lett.* **2014**, 112, 047203.

61. Tulapurkar, A. A.; Suzuki, Y.; Fukushima, A.; Kubota, H.; Maehara, H.; Tsunekawa, K.; Djayaprawira, D. D.; Watanabe, N.; Yuasa, S. Spin-torque diode effect in magnetic tunnel junctions. *Nature* **2005**, 438, 339–342.

62. Jiang, W.; Zhang, X.; Yu, G.; Zhang, W.; Jungfleisch, M. B.; Pearson, J. E.; Heinonen, O.; Wang, K. L.; Zhou, Y.; Hoffmann, A.; te Velthuis S. G. E. Direct Observation of the Skyrmion Hall Effect. arXiv:1603.07393 (accessed April 5, 2016).

63. Vansteenkiste, A.; Leliaert, J.; Dvornik, M.; Helsen, M.; Garcia-Sanchez, F.; van Waeyenberge, B. The design and verification of MuMax3. *AIP Adv.* **2014**, 4, 107133.

64. Thiaville, A.; Rohart, S.; Jué, E.; Cros, V.; Fert, A. Dynamics of Dzyaloshinskii domain walls in ultrathin magnetic films. *Europhys. Lett.* **2012**, 100, 57002.

65. Landau, L. D.; Lifshitz, E. M. On the theory of the dispersion of magnetic permeability in ferromagnetic bodies. *Phys. Z. Sowjetunion* **1935**, 8, 153–169.

66. Gilbert, T. L. A phenomenological theory of damping in ferromagnetic materials.

*IEEE Trans. Magn.* 2004, **40**, 3443–3449.

67. Slonczewski, J. Current-driven excitation of magnetic multilayers. *J. Magn. Magn. Mater.* **1996**, 159, L1–L7.

68. Fujita, N.; Inaba, N.; Kirino, F.; Igarashi, S.; Koike, K.; Kato, H. Damping constant of Co/Pt multilayer thin-film media. *J. Magn. Magn. Mater.* **2008**, 320, 3019–3022.

69. Shaw, J. M.; Nembach, H. T.; Silva, T. J. Resolving the controversy of a possible relationship between perpendicular magnetic anisotropy and the magnetic damping parameter. *Appl. Phys. Lett.* **2014**, 105, 062406.

**FIGURE CAPTIONS**

**Fig. 1**. **Layout showing the relationship between transverse forces ($F_g$ and $F_p$) and longitudinal drift velocity ($\mathbf{v}^d = v_x^d \hat{e}_x$) associated with an edge-meron.** The drift velocity of the meron depends on the driving current density (note that $\mathbf{j} = -J\hat{e}_x$). (**A**) For leftward injected electric currents, the Magnus force ($F_g$) reacts against the boundary's attraction force ($F_p$) and thus can result in steady-state meron motion for $J$ in a certain range. The force balance for a skyrmion, with a positive velocity, is shown for comparison. (**B**) For rightward flowing electric currents, the Magnus force then favors the attraction force, repelling the edge-meron out of the track.

**Fig. 2**. **Unidirectional motion of edge-merons.** (A–D) Steady-state drift motion of edge-merons under an in-plane current. (A) Snapshots of an edge-meron at indicated times. (C) Snapshots of an edge-meron as well as a skyrmion at indicated times. (B and D) The vertical component of normalized magnetization averaged over the entire volume of the nanotrack, $m_z$, and the topological charge, $Q$, as a function of the current action time, $t$, corresponding to (A and C), respectively. The track is 60 nm wide and 1 nm thick. $\alpha = 0.3$. In (A and B), $\beta$ is assumed to be $2\alpha$, namely, 0.6; the electric current is leftward injected as marked by the arrow and $J = +3.0 \times 10^{12}$ Am$^{-2}$. In (C and D), $\beta$ is set to be $0.5\alpha$, namely, 0.15; the applied electric current flows rightward as denoted by the arrow and $J = -5.0 \times 10^{12}$ Am$^{-2}$. From $F_g = (1-\beta/\alpha)Gv_x^s \propto (\beta/\alpha - 1)J$, it is clear that, for $\beta$ equal to twice and half of $\alpha$, $(1-\beta/\alpha)$ changes sign. Hence, the current directions must be reversed to maintain the fixed direction of the Magnus forces in (A and C). Moreover, the rigidity of the edge-meron is preserved in the motion process. The steady state is established a few nanoseconds after the application of the current, as revealed by the plateaus in the $m_z(t)$ and $Q(t)$ curves. (**E** and **F**) Destabilization and annihilation of an edge-meron under an in-plane current. (E) Snapshots of an edge-meron at indicated times. (F) $m_z$ and $Q$ versus $t$. Note that, here, all parameters in (E and F) are the same as in (A and B) except for the current direction. The current flows in the direction such that the Magnus force assists the drag force of the boundary. Under the joint forces, the edge-meron shrinks in size and loses the topological charge and finally is annihilated.

**Fig. 3. Phase diagram for transverse stabilization and longitudinal steady motion of the edge-meron subject to in-plane currents.** $\alpha=0.3$. In (**A**), $\beta=2\alpha$; each ring is centered at ($K_u$, $D$) with the inner and outer radii representing the lower and upper critical current densities; inside each ring, the steady-state meron motion is attainable. Apart from the radii, the colored peripheries in each ring also code the current densities. (**B** and **C**) The stabilization rings of an edge meron for different $\beta/\alpha$ values (all other parameters kept the same).

**Fig. 4. Mobility of edge-merons in steady drift motion.** (**A**) Drift velocity and (**B**) topological charge as a function of the current density. (**C₁–G₁**) The contours of spin configuration and (**C₂–G₂**) topological-charge density of the edge-meron under specified current densities. In (A), $\beta=2\alpha$ and $0.5\alpha$, with $\alpha=0.3$ and $0.01$, are considered for the edge-meron, and the skyrmion motion is checked for $\beta=2\alpha$ and $0.5\alpha$ with $\alpha=0.3$ for comparison. In (B), $\beta=2\alpha$ and $0.5\alpha$ are considered for the edge-meron only with $\alpha$=0.3. In both (A and B), several ($K_u$, $D$) combinations are taken into account to see the effect of the parameter variation. In (C–G), $K_u$=0.8 MJm$^{-3}$, $D$=3.5 mJm$^{-2}$, $\alpha$=0.3, and $\beta$=2$\alpha$. The results in (A) suggest that the edge-merons and skyrmions have the same mobility, exhibiting no dependency on the material parameters $K_u$ and $D$ and the topological charge $Q$, which is in qualitative agreement with the theoretical result [Eq. (2)].

**Fig. 5. Current-driven dynamics of an edge-meron in a notched nanotrack.** The triangular notch [inset in (A)] models boundary roughness in real samples. Here, the depth of the notch is 10% of the width of the track, namely, $h$=0.1$w$. $\alpha$=0.3 and $\beta$=2$\alpha$. In (**A** and **B**), the current density is $J$=+4.0×10$^{12}$ Am$^{-2}$ and +6.0×10$^{12}$ Am$^{-2}$, respectively. Specified in each subpanel is the action time of the electric current. Each central panel displays $m_z$ and $Q$ against the current action time.

**Fig. 6. Demonstration of a meron diode.** (**A**) Schematic architecture of the diode. The lateral junction made of a width-modulated nanotrack is the functional element, where the meron motions are modulated and the diode effect is realized. The 'carriers'—merons—are injected into the junction by a vertical current across a point-contact spin valve situated near the lower boundary. The junction is 'biased' by an in-plane current supplied by the control circuit; once created, a meron goes into one of the two motional modes, depending on the current direction. The detection circuit outputs a signal by recording the magnetic-flux variation across the coils attached atop the narrow arms of the junction. Note that, here, the out-of-plane and in-plane current densities are denoted as $J_1$ and $J_2$, respectively. (**B**–**D**) The forward and (**E**–**G**) reverse operations on the diode. Each instance contains 6 operation cycles. (B and E) Current sequences used to inject and manipulate the edge-merons. (C and F) Evolution of $m_z$ with the operational time. (D and G) Carrier distribution inside the junction at characteristic times. (**H**) '*I*–*V*' (here, $\varepsilon$–*J* indeed) curve of the meron diode. Here, $\varepsilon$ is the induced electromotive force in the coils. (**I**) Domain-wall width as a function of in-plane current density *J* for forward cycle. The domain-wall pair is converted from the meron at the connection area of the wide and narrow arms. Here, $w_b=3w_n=60$ nm and $l_b=l_n=200$ nm.

**Figure 1**

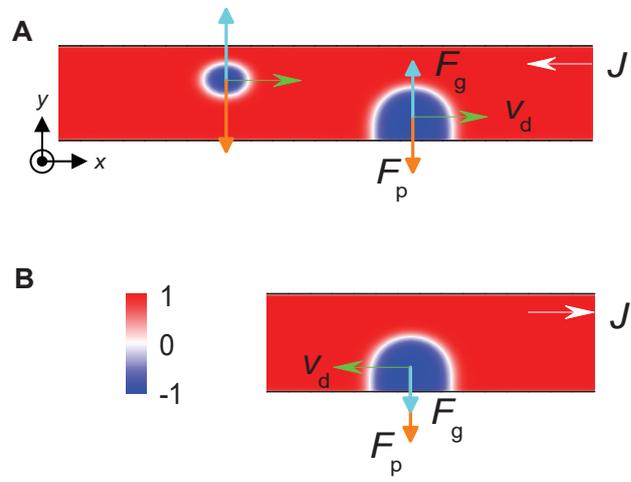

**Figure 2**

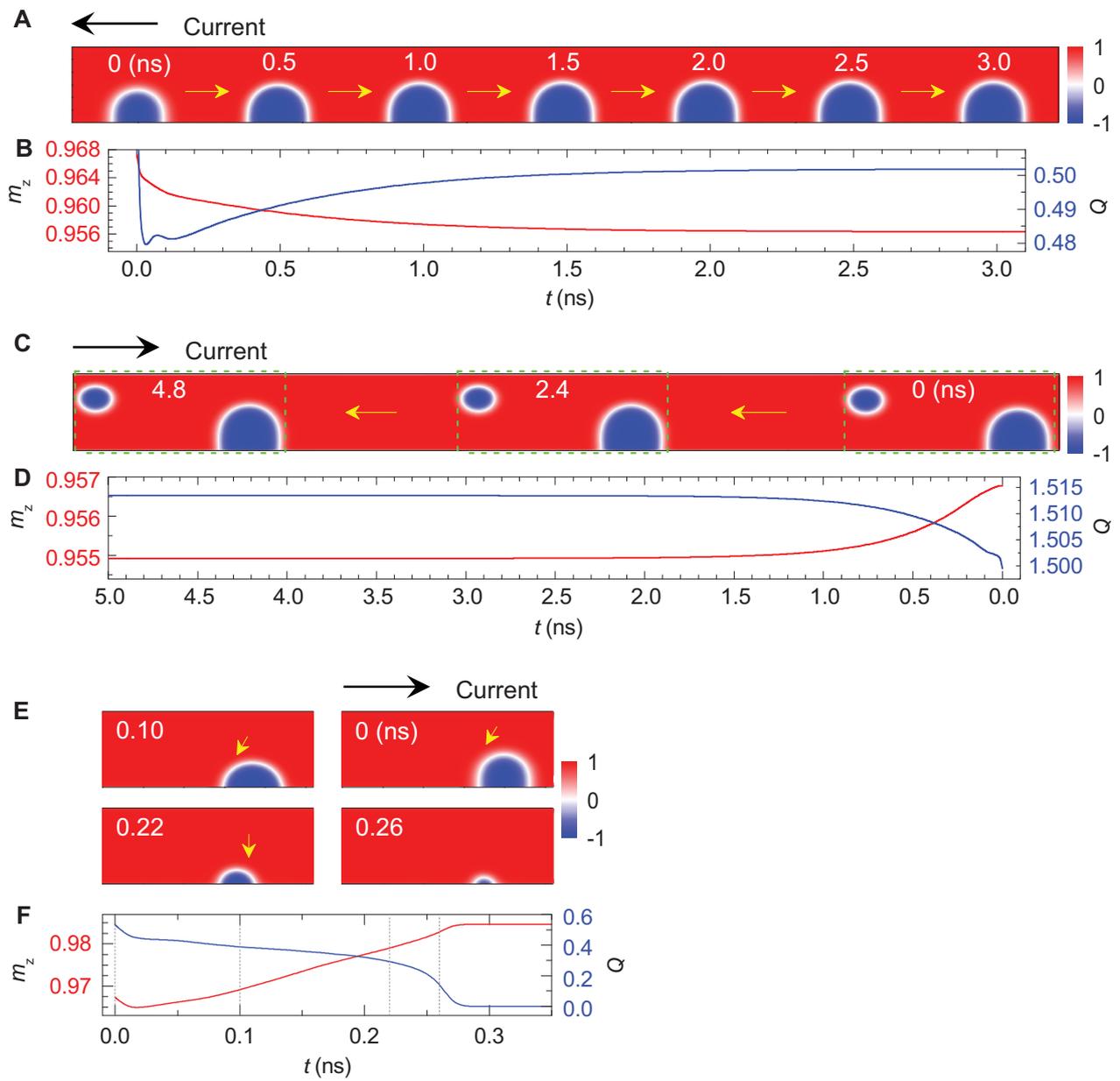

**Figure 3**

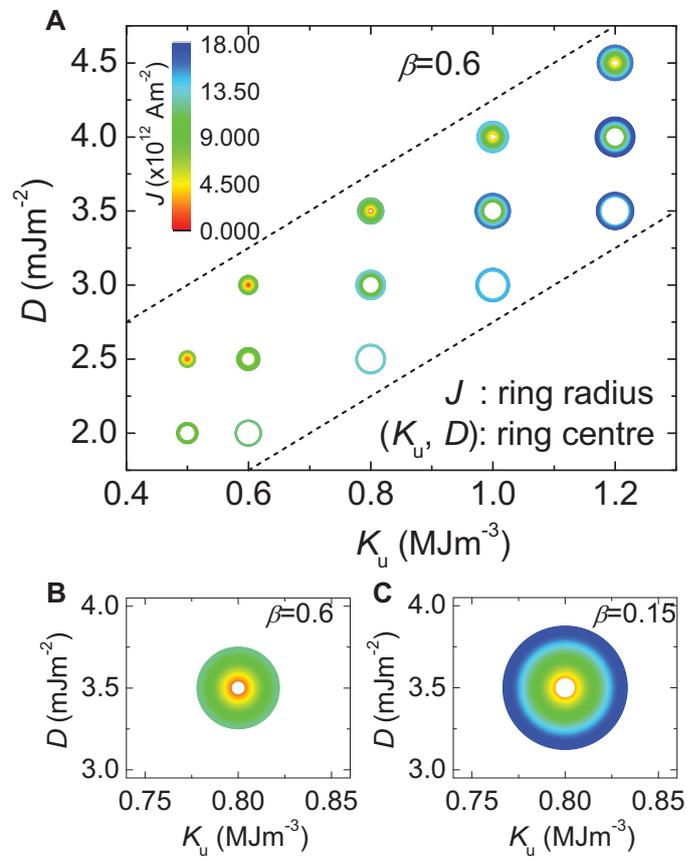

**Figure 4**

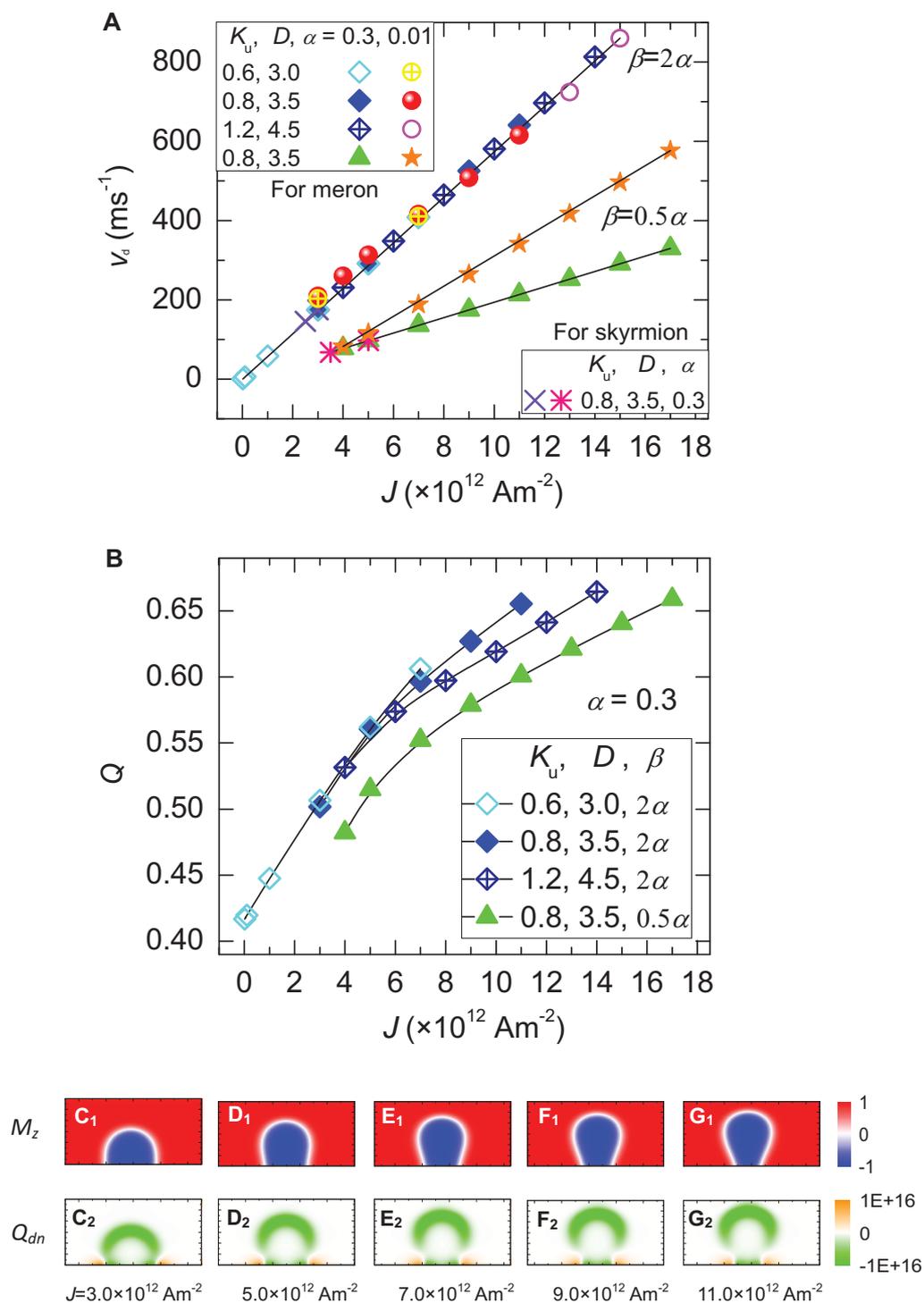

**Figure 5**

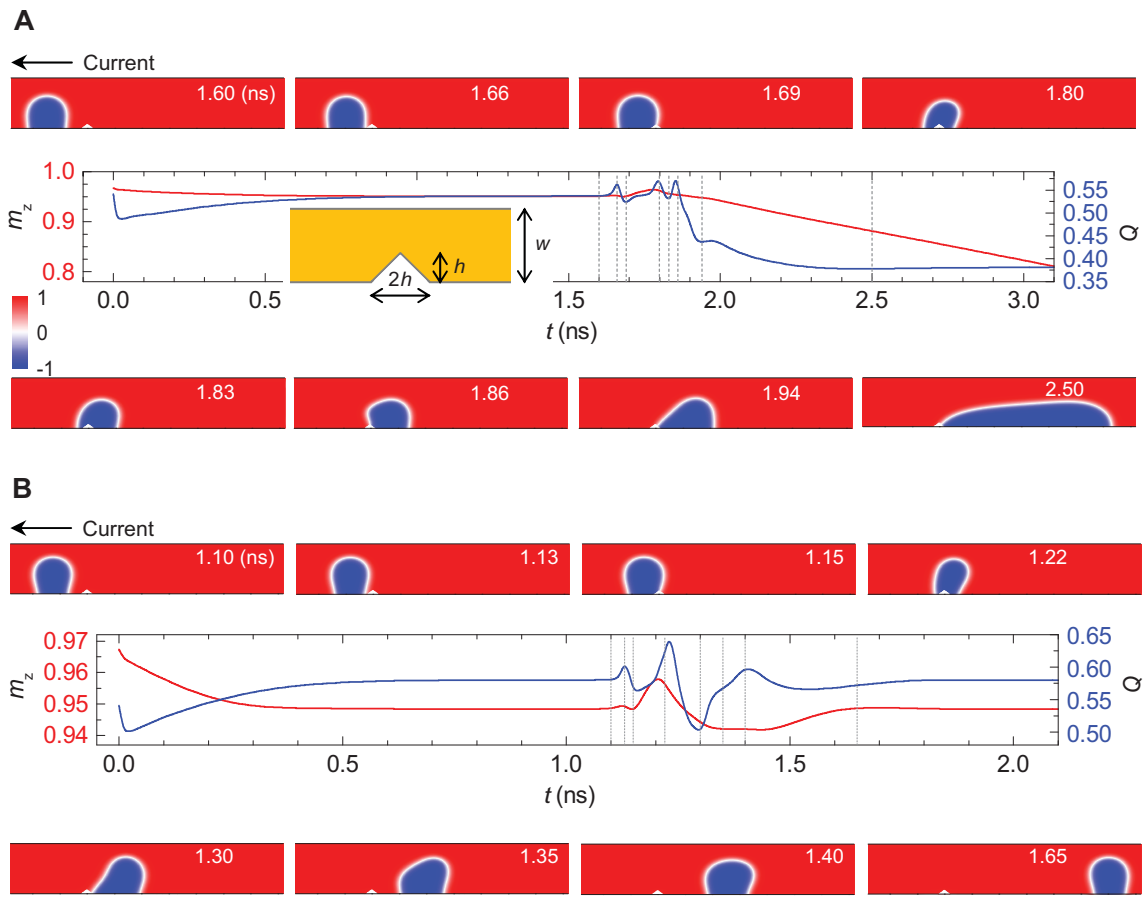

# Figure 6

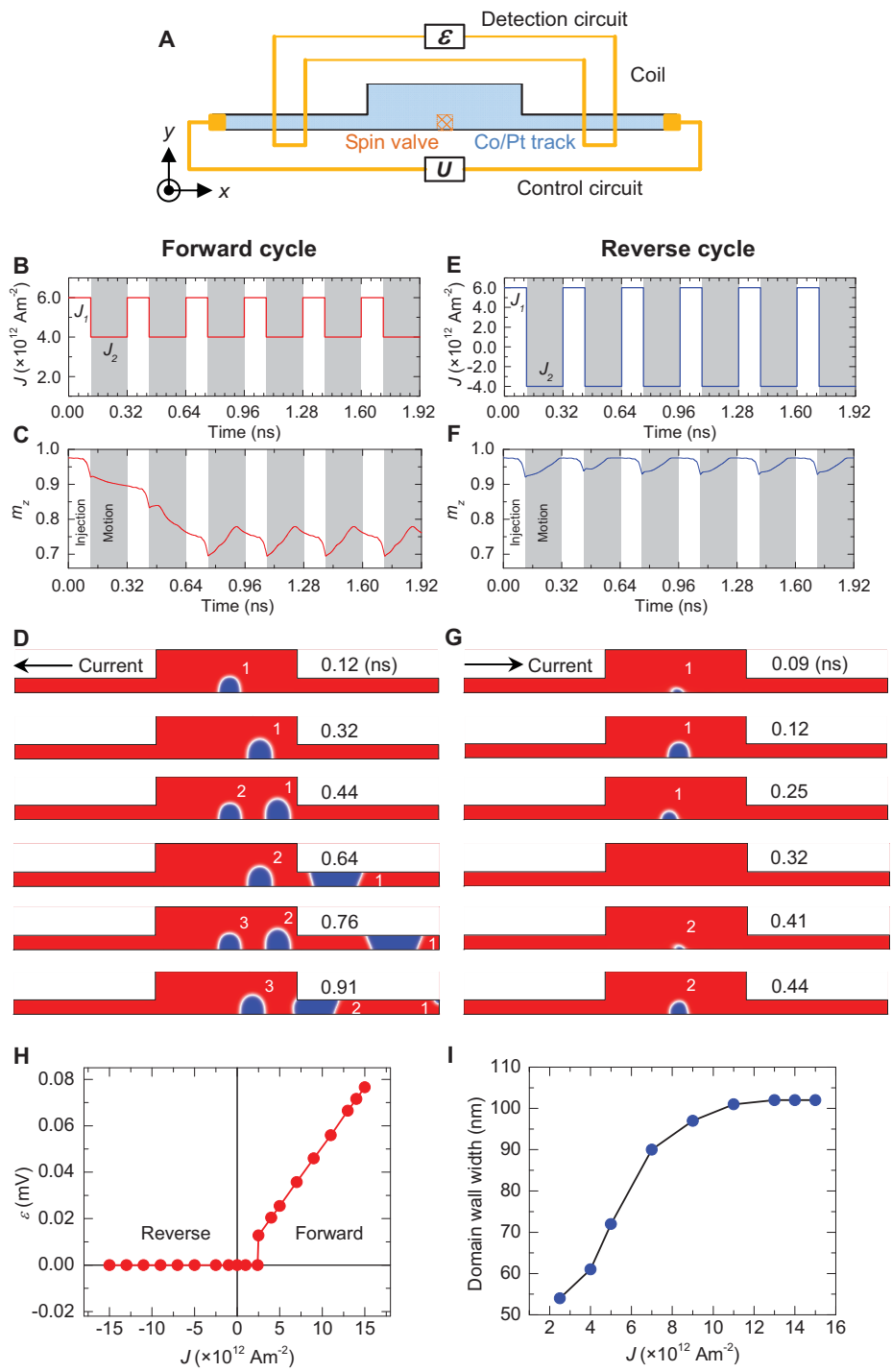